*Research*

# The use of social media among library professionals and patrons: A review of literature

[1]Abimbola Labake Agboke* and [2]Felicia Nkatv Undie

[1]University Library, University of Uyo, Uyo. Akwa Ibom State, Nigeria.
*Corresponding author's E-mail: E-mail: bimbo.agboke@yahoo.com
[2]Department of Library and Information Science, Cross River University of Technology Calabar, Nigeria.
E – mail: nkatvfelicia@gmail.com

This paper focused on the utilization of social media by the library professionals and library users. It provides the understanding of social media, the most popular social media platforms utilized in the libraries. It also mentions the reasons for adoption of social media in libraries be it academic, public, school libraries and other types of library. This is a review paper on the use of social media among library professionals and patrons. The findings reveal the contributions of social media to the libraries. Social media makes things easy for the library professionals and library users. It enables them to connect, create awareness to new information, to disseminate information instantly and helps to market the library resources and services. Therefore, it is recommended amongst others that the library management board should encourage the use of social media in libraries.

## INTRODUCTION

The use of social media has been of great interest to researchers in economics, psychology, and other fields like library and information science. Currently, the use of Internet communication tools is increasing rapidly in the libraries. The duties of library professionals are to acquire resources, process them and make them available to the library users or patrons. The introduction of media technology into the libraries, has changed the libraries' daily routines to technological way of working. The libraries' daily routines like acquisition, cataloguing, circulation of resources and other functions of library are done electronically. According to International Federation of Library Associations (IFLA), (2013). Libraries are not piles of books anymore; the general library environment has changed from analogue to digital. Library automation systems have helped libraries to provide easy access to their collections through the use of computerized library catalogues (On-line Public Access Catalog – OPAC) which more recently led to digital libraries. Most of the resources in the libraries are now digitized and can be accessed through Web that is library 2.0 or social media.

According to Thurairaj et al. (2015), social networking sites are defined as mobile- or Internet-based social platforms created and designed to enable users to communicate, collaborate and share content across contacts and communities. With the introduction of Information and communication Technology (ICT) in library fields, the adoption of social media has been widespread to keep the library relevant by utilizing the possibilities of the Web. Social media is an instrument of communication. Social media includes networking web sites like Facebook, MySpace, microblogging web sites like Twitter and other media like blogs, podcasts, photos and videos. By posting library material via social media on library page, it can be used by the variety of locations on the Web (Tuten, 2001).

Kaplan and Haenlain (2010: 61) defined social media "as a group of Internet-based applications that build on the ideological and technological foundations of Web 2.0, and that allow the creation and exchange of User Generated Content". They further clarify the relationship between social media, Web 2.0, and "user generated content". Web 2.0 is seen as the ideological and technological foundation of social media. User generated content is, in turn, all the ways in which people use social media. Web 2.0 is according to this definition a broader concept than social media. Social media is, however, also often used as a synonym for Web 2.0 (Anttiroiko and Savolainen, 2011). The important features of social media sites where these technologies are at play are the users' central role, the ability to form connections between users, and the ability to post content in different forms (Anderson, 2007; Cormode and Krishnamurthy, 2008). The advent of social media in libraries is a shift in paradigm from analogue to computer and smartphones and utilization of library resources could take place without physically present in the library. The aim of social media in library is to keep the library relevant by supporting utilization of the library through Web. This study is important because it will enable the researchers to know the popular social media networking sites utilized by library and reasons for their adoption by library professionals and library patrons. This review will serve as a database for researcher writing on social media.

**Popular Social media networking sites utilized by the library**

Many researchers have researched into the popular social media used in the libraries. Ezeani and Igwesi (2012), stated that some popularly used social media networking sites by libraries in Nigeria to meet the information needs of users are Facebook, MySpace, Ning, Blogs, Wikis, LinkedIn, Twitter, YouTube, Flickr and Library Thing. They enable the library professionals and library users to collaborate and connect for information utilization. Olajide et al. (2017), reported that Facebook enjoy the highest patronage among the Nigerian libraries with all the 24 libraries used in the study using Facebook, followed by yahoo with 7 libraries and then google+ with 6 libraries. The least patronized social media platforms were 2go and YouTube with just 1 library using them followed by WhatsApp which is used by 2 libraries while Skype and others were being used by 3 libraries.

According to Chao and Keung (2017), Facebook is one of the most popular social networking sites worldwide, with 1.71 billion active users in 2016. Anwar and Zhiwei (2019), opined that a number of social networks have been launched and some of them are very much popular throughout the globe like Facebook, Twitter, YouTube, WeChat, Instagram, QQ, QZone, Weibo, Twitter, Tumblr, Telegram, Baidu Tieba, LinkedIn, LINE, Snapchat, Pinterest, Viber etc. World Wide Worx and Fuseware (2016), stated that the most popular social networking sites include Facebook, Twitter, Instagram, you tube, Snapchat and WhatsApp. Ezeani and Eke (2011), reported that the most applicable web 2.0 technology for library services is the social networking tools – where librarians can interact with their users to study their needs and give a feedback; photo sharing – where archival pictures can be posted to users or uploaded on the library websites…" Librarians in Nigeria are gradually utilizing these tools to offer "on the spot" library services to users. The use and development of various social media services continue to increase across the world because of the value they have in aiding human communication.

**Reasons for the adoption of Social Media Network Sites among the Librarians and the Library Patrons**

*For information dissemination*

Social media provides real-time information dissemination among the library professionals and library patrons. According to Anwar and Zhiwei (2019), social media helps the library professionals to make things easy for them and for their readers to increase their capacity to build good relationships among library staff and library users. Social media enables the library professional bodies like Nigerian Library Association (NLA), West African Library Association (WALA) in Africa and others to interact through internet communication. The National Library of Australia (NLA), (2010) stated that Library supports every employee to have the opportunity to communicate online via social media professional networking sites, blogs and personal web sites. The National Library of Australia uses a variety of social media for notifying news, relevant items from collection and library events. Facebook is used by the library to inform the library users about major events, activities and recent acquisitions through posting photos, videos and links to resources about the library. Library uses YouTube for sharing videos of many of the events held at the library. Presentations and talks organized on different topics by the library are disseminated via podcasts (NLA, 2010).

Furthermore, Chu and Du (2013) stated that social media is putting a massive impact on libraries and information centers to promote library services and sources. Several social media bring all the library users community together on one spot to share their ideas and views about their relevant and specific information. Also, social media is providing massive space to the library professionals to create a virtual environment to enhance the library service providing capacity. Taylor and Francis (2017), asserted that the use of social media is making things easy for library professionals to reduce the gap between library users and library resources and services.

Moreover, social media facilitates library professionals to achieve their academics goals and objectives. In Nigeria, Olajide et al. (2017) revealed that the questionnaire of their research work was sent out to librarians through the Nigerian Library Association (NLA) online forum email and the Facebook page. A reminder message was sent every month on the online forum and the Facebook to achieve good response.

Social media enables the library professionals to create an account to promote their library sources and services. Mahmood and Richardson (2011), found that Really Simple Syndication (RSS) was the most popular Web 2.0 application in US academic libraries, for publishing news, sharing items published on library blogs, providing information on literacy instruction, information on new acquisitions, podcasts, vodcasts, databases and e journals. Other tools used included instant messaging (IM) (95%), social networking sites (SNS) (87%), blogs (85%), micro blogs (82%), social bookmarking (55%), whilst 47% of the libraries used Flicker for sharing photos of events.

### *For provision of current awareness services*

Social media creates an awareness of arrival of new information resources to the users without necessarily visiting the library physically. Gao et al. (2012), reviewed literature about Twitter published between 2008 and 2011. They concluded that Twitter allowed students to participate with each other in class (by creating an informal "back channel"), and extend discussion outside of class time. They also reported that students used Twitter to get up-to-date news and connect with professionals in their field. Kho's paper (2011), which explores social media use for customer engagement, substantiates the successful use of YouTube to enable users to embed videos in other Web 2.0 tools, such as Facebook, blogs and wikis. Flickr is a photo sharing website which allows users to store, sort, search and post photographs and to create discussion groups. Si et al. (2011), comment that tools such as wikis and RSS feeds are mostly used for searching for information and following current events respectively.

Through social media, instant interaction is made possible. Makori, (2011:34) emphasized that instant interaction between a librarian and the user is one of the many reasons academic libraries have chosen to integrate social media tools into their daily work. Social media tools such as YouTube, MySpace, Facebook, Twitter and Flickr are popular for online collaboration, communication and sharing of information among librarians and instant messaging (IM) provides the basis for librarians to interact with their patrons (Munatsi 2010: 255). Social media encourages collaborations among the librarians. Librarians are now able to communicate instantly with users by remotely providing assistance such as virtual reference services (Stephens 2006) and to offer current information to students and researchers (Xu et al. 2009). That is, instead of arranging face-to-face orientation programs, libraries are increasingly using RSS feed readers, podcasts and recorded videos to deliver audio and video commentary and instruction to users remotely.

### *For Marketing and promotion of information resources and services available in the libraries*

Libraries are users oriented, be it academic libraries, special libraries, school libraries and others and the aim of any library is for users' satisfaction. Libraries can achieve this through the use of social media. This is in line with the view of Olajide et al. (2017) that social media's presence is almost everywhere and the vision of the library within the last few years has been that library services should go to users not necessarily that users should come physically to the library. Therefore, libraries adopted social media has marketing and publicity tools in the libraries. Hendrix et al. (2009), stated that libraries are using Facebook mainly to market the library, push out announcement to library users, post photos, provide chat reference and have a presence in the social network. In another work by Rogers (2012), on web 2.0 application used by libraries to promote and market services, the librarians' response showed that 70.7% use blog, 66.7% use social networking sites, and instant messaging was 50.7% as the most commonly used ones.

In developed countries libraries are using latest trends to market their services. United States (U.S.) libraries of all types are increasingly using social media tools and Web 2.0 applications to connect with library users and to make library programs and services accessible (American Library Association ALA, 2001). Library of Congress is also utilizing social media for marketing its services and to interact with its online users. Library uses Blogging, Flickr, YouTube, Social Networking, iTunes and Twitter for its different services (Braziel, 2009). Harinarayana and Raju (2010), reiterated that social media technologies are becoming increasingly popular for use in academic libraries' marketing strategies. Social media tools have assisted university libraries in the promotion of information services to their patrons.

Kronqvist – Berg (2014), Studied the implementation of social media into the libraries. She reported that the main reason for introducing Library 2.0 services was, to develop the library (37.8% of n=82) and is especially supported by library managers. This option was followed by to market the library (22.0%) and to keep the library relevant (22.0%). The option with least support among the respondents was to give the library a modern impression (13.4%). Avid social media users saw the implementation of Library 2.0 mainly as a way to develop the library, followed by marketing possibilities, and keeping the library relevant. Social media enable communication, instant messaging, connection, information sharing and interactions between the librarians, library professional bodies and library users.

**CONCLUSION / RECOMMENDATIONS**

Social media has contributed to the development of libraries in various ways. It makes the library relevant in this digital age. Social media is used as marketing and promotional tools to get the users informed of the available resources and services in the library. The library professionals' bodies connect and collaborate through the use of social media. Library professionals disseminate information to the users through social media network. They can post recent book available in the library through Facebook, Twitter, blog and others to users. Date dues can be posted through WhatsApp, LinkedIn, Twitter and others. Inter library loans services is made possible through social media sites like Flickr and Library Thing.

Considering the numerous functions of social media in library practices, the following recommendations were made: All libraries should adopt the use of social media in their daily activities. Furthermore, Librarians should create social media platforms for their users to enable them communicate with the library users or patrons. Also, the Library management board should make provision for free browsing systems for all the users and library professionals. More desk top computers and laptops should be provided into the libraries by the library management. Workshops and seminar on the use of social media should be conducted for the library professionals and patrons.

**REFERENCES**

American Library Association (ALA) (2001). Libraries making good use of social media and Web 2.0 applications. Retrieved from *http://www.ala.org/news/mediapresscenter/americaslibraries/socialnetworking*

Anderson, P. (2007). What is web 2.0? Ideas, technologies and implications for education. JISC Technology & Standards Watch. London: JISC. Retrieved from *http://www.jisc.ac.uk/media/documents/techwatch/tsw0701b.pdf*

Anttiroiko, A.V., & Savolainen, R. (2011). Towards Library 2.0: The adoption of Web 2.0 technologies in public libraries. Libri, 61(2): 87-99. doi: 10.1515/libr.2011.008.

Anwar, M. & Zhiwei, T. (2019). Social media makes things possible for Librarians: A Critical Note. Amer. Jour. of Biomed. Sci. & Res. 6(1) AJBSR.MS.ID.000985. DOI: *10.34297/AJBSR.2019.06.000985.*

Braziel, L. (2009). *Social media marketing example #12: Library of Congress*. Retrieved from http://www.ignitesocialmedia.com/social-media-examples/social-media-marketing-example-library-of-congress

Chao, C. & Keung, N. (2017). Predicting social capital on Facebook: The implications of use intensity, perceived content desirability, and Facebook-enabled communication practices'. Comp. in Human Behav. 72: 259–268. dl.acm.org/doi/10.1016/j.chb.2017.02.058.

Chu, S.K.W & Du, H. S. (2013). Social networking tools for academic libraries. Jour. of librnship and info. Sci. 45(1): 64-75.

Cormode, G., & Krishnamurthy, B. (2008). Key differences between Web 1.0 and Web 2.0. *First Monday, 13*(6). doi:10.5210%2Ffm.v13i6.2125.

Ezeani, C. N, & Igwesi.U. (2012). Using social media for dynamic library service delivery: The Nigeria experience. Libr. Phil. and Prac. (e-journal). 814. *https://digitalcommons.unl.edu/libphilprac/814*

Ezeani, C. N., & Eke, H. N. (2011). Transformation of Web 2.0 into Lib 2.0 for driving access to knowledge by academic libraries in Nigeria. In the 48th National Conference and Annual General meeting of the Nigerian Library Association Theme: *Knowledge management for national development.* Ibadan: HEBN Publishers, p.80

Gao, F., Luo, T., & Zhang, K. (2012). Tweeting for Learning: A Critical Analysis of Research on Microblogging in Education Published in 2008-2011. British Jour. of Educational Tech. 43: 783-801. *http://dx.doi.org/10.1111/j.1467-8535.2012.01357.x*

Harinarayana, N. & Raju V. (2010). Web 2.0 features in university library web sites. The Electronic Library, 28(3): 69-88.

Hendrix, D; Chiarella, C; Hasman, L; Murphy, S. & Zafron, M.L. (2009). Use of Facebook in Academic Health Sciences Libraries. Jour. of the Med. Libr. Assn. 9 (1): 44-47.

International Federation of Library Associations and Institutions (IFLA). (2013). Bridging the Digital Divide: making the world's cultural and scientific heritage accessible to all, [online] IFLA/UNESCO Manifesto for Digital Libraries. Available at: http://ifla.org

Kaplan, A. M. & Haenlein, M. (2010). Users of the world, unite! The challenges and opportunities of social media. Business Horizons, 53(1): 59-68. doi:10.1016/j.bushor.2009.09.003

Kho, N. D. (2011). Social media in libraries: keys to deeper engagement. *Information Today, 28(6), 1, 31-32.*

Kronqvist – Berg, M. (2014). Social media and public libraries: Exploring Information activities of library professionals and users. Abo: Åbo Akademi University Press.

Mahmood, K. & Richardson, J. (2011). Adoption of Web 2.0 in US academic libraries: a survey of ARL library websites Program. 45 (4): PP.365 – 375.

Makori, E.O. (2011). Bridging the information gap with the patrons in university libraries in Africa: the case for investments in Web 2.0 systems. *Library Review,* 61(4): 340-350.

Munatsi, R. 2010. Implementation of library 2.0 services in African academic and research libraries: need for fundamental rethink. [Online]. http://www.scecsal.org/conferences/2010/spapers2010.pdf.

National Library of Australia (NLA). (2010). *National Library of Australia publishes social media guidelines.* http://web.resourceshelf.com/go/resourceblog/62891

Olajide, A. A, Otunla, A.O. & Omotayo, O.A. (2017). How libraries are using social media:

Nigeria perspective. International Jour. of Digital Libr. Ser. 7 (3): 79 – 94. IJODLS | Geetanjali Res. Pub. www.ijodls.in   ISSN:2250-1142 (Online),

Rogers, C.R. (2012). Social media, libraries and web 2.0: How American libraries are using new tools for public relations to attract New Users- Fourth annual survey November 2011. Available at dc.statelibrary.sc.gov/bitstream/handle/10827/7271/SCSL_Social_Media_Libraries_2011.pdf?sequence=1&isAllowed.

Si, L., Shi, R. & Chen, B. (2011). An investigation and analysis of the application of Web 2.0 in Chinese university libraries. The Elect. Libr., 29(5): 651-668.

Stephens, M. (2006). Exploring Web 2.0 and libraries. Libr. Tech. Rep. 42(6): 8-14.

Taylor & Francis (2017). How libraries are applying social media. Content & Communications Team, USA.

Thurairaj, S., Hoon, E.P., Roy, S.S., Fong, P.W., Tunku, U., Rahman, A. et al., (2015). 'Reflections of students' language usage in social networking sites: Making or marring academic English', *University Tunku Abdul Rahman* 13(4):302–316.

Tuten, T.L. (2001*). Advertising 2.0 - Social media marketing in a Web 2.0 World. Book review.*
*Retrieved May 9, 2012, from http://blogcritics.org/books/article/book-review-advertising- 20-social-media/*

World Wide Worx & Fuseware, (2016). South African social media landscape 2016, Retrieved 09 May 2017, from http://www.worldwideworx.com/wp-content/uploads/ 2016/11/Exec-Summary-Social-Media-2016.pdf.

Xu, C., Ouyang, F. and Chu, H. (2009). The academic library meets Web 2.0: applications and implications. Jour. of Academic Libranship, 35(4): 324-331.